%% file: main.tex
\documentclass[onecolumn,numbers]{els-mrw} 

\usepackage{amsmath,amssymb,amsfonts,amsthm,makeidx,graphicx}
\usepackage{txfonts}
\usepackage{helvet}
\usepackage{xspace}
\usepackage[normalem]{ulem}
\usepackage{xcolor}
\definecolor{linkorange}{RGB}{230,159,0}
\definecolor{linkblue}{RGB}{86,180,233}
\usepackage[colorlinks=true,breaklinks=true]{hyperref}
\AtBeginDocument{\hypersetup{citecolor=linkorange,linkcolor=linkorange,urlcolor=linkblue}}

\newenvironment{keypoints}{\bgroup%
\par\addvspace{\baselineskip}%
\abstractfont\leftskip9mm\rightskip9mm\noindent{\parindent=0pt\abstractheadfont\textbf{Key Points:}}\\[-0.1cm]%
\begin{itemize}\itemsep0pt\parsep0pt\topsep0pt}{\end{itemize}\egroup}

\graphicspath{{figs/}}

\begin{document}


\input{sections/notations}

\chapter{Scattering Theory}\label{chap1}

\author[1]{Christoph Hanhart}%
\author[2]{Mikhail Mikhasenko}%

\address[1]{\orgname{Forschungszentrum J\"ulich}, \orgdiv{Institute for Advanced Simulation (IAS-4)}, \orgaddress{D-52425 J\"ulich, Germany}}
\address[2]{\orgname{Ruhr-Universit\"at Bochum}, \orgdiv{Institut f\"ur Experimentalphysik}, \orgaddress{D-44780 Bochum, Germany}}

\renewcommand{\articlecopyright}{}
\articletag{}

\maketitle

\begin{abstract}[Abstract]
    This chapter provides an overview of the fundamental framework of scattering theory, which is widely used in particle physics to describe and interpret interactions among elementary particles. We explore how transition amplitudes enable the analysis of data from scattering experiments and
    the extraction of resonance parameters.
    The concepts and methods discussed are essential for understanding processes studied at major accelerator facilities worldwide and in a broad range of hadronic and nuclear systems.
\end{abstract}

\begin{keywords}
    $S$-matrix \sep unitarity \sep analyticity \sep scattering amplitude \sep resonances \sep partial-wave decomposition \sep cross section \sep decay rate \sep Riemann sheets
\end{keywords}

\begin{keypoints}
    \item Develop the $S$-matrix formalism as the bridge between theoretical amplitudes and measurable quantities such as cross sections and decay rates
    \item Demonstrate how partial-wave decomposition enables systematic analysis of resonances with specific angular momentum properties
    \item Explore analytic continuation techniques for characterizing resonance phenomena across multiple Riemann sheets
    \item Showcase applications to modern experimental analysis, demonstrating theoretical amplitude construction and useful parameterization
\end{keypoints}

\begin{glossary}[Nomenclature]
    \begin{tabular}{@{}lp{34pc}@{}}
        CM   & Center-of-momentum (frame)                      \\
        EFT  & Effective Field Theory                          \\
        BW   & Breit-Wigner (amplitude)                        \\
        BS   & Bethe-Salpeter (equation)                       \\
        LSZ  & Lehmann-Symanzik-Zimmermann (reduction formula) \\
        PDG  & Particle Data Group                             \\
        QFT  & Quantum Field Theory                            \\
        UV   & Ultraviolet (behavior/limit)                    \\
        b.g. & background                                      \\
        thr. & threshold                                       \\
    \end{tabular}
\end{glossary}

\section{Introduction}
\label{sec:introduction}

The study of fundamental interactions in particle physics relies heavily on our ability to describe and interpret the outcomes of particle collisions. From the early discoveries of hadronic resonances to modern precision measurements at the Large Hadron Collider and other experiments around the world, understanding how particles interact and transform has been central to advancing our knowledge of the fundamental forces of nature.
The main experimental tools used to study particle, hadron or nuclear physics are elastic (where the initial and final states are the same) or inelastic scattering (where the initial and final states are different). In these experiments, the particles leaving the interaction region are identified and their properties are measured. In exclusive reactions, for example, the number of particles of a given kind flying in a certain direction with a given momentum is counted. Scattering theory is then used to relate these rates to scattering amplitudes, which can be interpreted theoretically. If in these reactions there are narrow (long-lived) intermediate states, it is often convenient to start the theoretical analysis from the decay of that state---then we speak of a decay amplitude. Decay and scattering amplitudes are both constrained through the conservation of probability or, mathematically speaking, the unitarity of the scattering matrix. This framework also relies on analyticity arising from causality, and partial-wave decomposition enabling systematic studies of resonances with definite angular momentum properties. Modern applications of scattering theory extend far beyond traditional particle physics, finding relevance in nuclear physics, condensed matter systems, and even cosmological studies.

The chapter is structured as follows: after some generalities in the next section, where the $S$-matrix as well as typical observables are introduced, we discuss the partial-wave decomposition in Sec.~\ref{sec:partial-waves}. In Sec.~\ref{sec:analyticity} and \ref{sec:unitarity} the concepts of analyticity and unitarity are introduced, followed by a discussion on analytic continuation and resonance properties in Sec.~\ref{sec:continuation}, and finally in Sec.~\ref{sec:theoretical} we discuss how scattering amplitudes are calculated conceptually as well as a concrete parameterization. The chapter closes with a short summary.

\input{sections/ScatteringTheory}\label{sec1:subsec1}

\section{Conclusions}
\label{sec:conclusions}

This chapter presents the $S$-matrix formalism as the fundamental framework connecting quantum field theory to experimental observables in particle and hadron physics. We have shown how transition amplitudes determine measurable quantities such as cross sections and decay rates, thereby linking the underlying dynamics to high-energy collision data. The discussion highlighted three essential principles: partial-wave decomposition for systematic resonance analysis, analyticity constraints arising from causality, and unitarity ensuring probability conservation. The power of analytic continuation to multiple Riemann sheets was illustrated for characterizing complex resonance phenomena and multi-channel processes. These theoretical foundations, combined with modern computational methods and effective field theory approaches, provide the essential tools for interpreting current experimental results and guiding future discoveries in fundamental particle interactions.

\begin{ack}[Acknowledgments]%
    We thank St\'ephane T'Jampens for valuable comments on the manuscript.
\end{ack}



\section*{Declaration of AI-assisted technologies in the writing process}

During the preparation of this work the authors used GitHub Copilot, Cursor with GPT-4.1, 4.2, and 5.3 models in order to polish phrasing in the text and perform orthographical checks. After using these tools, the authors reviewed and edited the content as needed and take full responsibility for the content of the publication.

\bibliographystyle{elsarticle-num}
\bibliography{reference}

\end{document}

%% file: sections/notations.tex
\newcommand{\pc}{P_{c\bar c}}
\newcommand{\jp}{J/\psi p}
\newcommand{\sigh}{\Sigma_c^{(*)}\bar{D}^{(*)}}
\newcommand{\sigd}{\Sigma_c\bar{D}}
\newcommand{\sigdstar}{\Sigma_c\bar{D}^*}
\newcommand{\sigstard}{\Sigma_c^*\bar{D}}
\newcommand{\lamh}{\Lambda_c\bar{D}^{(*)}}
\newcommand{\etacp}{\eta_c p}
\newcommand{\etacn}{\eta_c p}
\newcommand{\dsz}{D_{s0}^*(2317)}
\newcommand{\dsone}{D_{s1}(2460)}
\newcommand{\dz}{D_0^*(2300)}
\newcommand{\done}{D_1(2430)}
\newcommand{\bdpp}{B^-\to D^+\pi^-\pi^-}
\newcommand{\cccc}{cc\bar c\bar c}
\newcommand{\bra}[1]{\left\langle #1 \right|}
\newcommand{\ket}[1]{\left| #1 \right\rangle}
\newcommand{\braket}[2]{\left\langle #1 | #2 \right\rangle}
\newcommand{\sclen}{a}
\newcommand{\rhoinel}{\rho_\text{inel.}}

\newcommand{\Br}{\textrm{Br}}
\newcommand{\Disc}{\textrm{Disc}}
\newcommand{\Imag}{\textrm{Im}}
\newcommand{\BW}{\textrm{BW}}
\newcommand{\Chew}{\varSigma}
\newcommand{\thr}{\textrm{thr.}}
\newcommand{\bg}{\textrm{b.g.}}
\newcommand{\res}{\textrm{R}}

\newcommand{\diff}{\mathrm{d}}

\newcommand{\Rcal}{\mathcal{R}} 

\newcommand{\Kcal}{\ensuremath{\mathcal{K}}\xspace} 
\newcommand{\Mcal}{\ensuremath{\mathcal{M}}\xspace} 
\newcommand{\Vcal}{\ensuremath{\mathcal{V}}\xspace} 
\newcommand{\Gcal}{\ensuremath{\mathcal{G}}\xspace} 
\newcommand{\Acal}{\ensuremath{\mathcal{A}}\xspace} 
\newcommand{\Ncal}{\ensuremath{\mathcal{N}}\xspace} 
\newcommand{\Pcal}{\ensuremath{\mathcal{P}}\xspace} 
\newcommand{\Qcal}{\ensuremath{\mathcal{Q}}\xspace} 
\newcommand{\Bcal}{\ensuremath{\mathcal{B}}\xspace} 

\newcommand{\fin}{\text{fin}}
\newcommand{\ini}{\text{ini}}
\newcommand{\finini}{\text{f;i}}
\newcommand{\iepsilon}{i\epsilon} 

\newcommand{\MI}{\ensuremath{\Mcal_{I}}\xspace}
\newcommand{\MII}{\ensuremath{\Mcal_{II}}\xspace}
\newcommand{\MIII}{\ensuremath{\Mcal_{III}}\xspace}
\newcommand{\MIandII}{\ensuremath{\Mcal_{I,II}}\xspace}

%% file: sections/ScatteringTheory.tex


\section{Generalities}\label{sec:generalities}
\input{sections/generalities}

\section{Partial waves}\label{sec:partial-waves}
\input{sections/partial-waves}

\section{Analyticity}\label{sec:analyticity}
\input{sections/analyticity}

\section{Unitarity}\label{sec:unitarity}
\input{sections/unitarity}

\section{Analytic continuation}\label{sec:continuation}
\input{sections/continuation}
\section{Theoretical description}\label{sec:theoretical}
\input{sections/theoretical}



%% file: sections/generalities.tex
This section establishes the mathematical foundations of scattering theory, beginning with the fundamental concepts and building toward the practical tools needed for analyzing experimental data.

In particle, hadron, and nuclear physics, the transition amplitude serves as a fundamental mathematical object that quantifies the quantum-mechanical probability for a system of initial particles to evolve into a well-defined final state. The squared magnitude of the amplitude, when integrated over appropriate phase-space variables and normalized by flux factors, determines measurable quantities such as cross sections and decay rates. Two primary kinematics govern experimental investigations: \textit{scattering processes}, where two initial particles collide, and \textit{decays}, where a single unstable particle transitions into multiple final states. The term scattering amplitude is synonymous with \textit{matrix element}---a reference to its origin in the \textit{S-matrix (scattering matrix)} formalism, where it represents the transition amplitude between initial and final states in the Hilbert space of quantum states.

Scattering amplitudes are primarily employed to describe \textit{exclusive processes}, where both the initial and final states are fully specified (e.g., $e^+ e^- \to \mu^+ \mu^-$). In contrast, \textit{inclusive processes} involve sums over classes of final states (e.g., $B^+ \to J/\psi + X$, where $X$ denotes any undetected particles), often due to limited experimental resolution or theoretical interest in aggregated outcomes. For inclusive measurements, observables depend not on individual amplitudes but on the summed probabilities derived from all contributing exclusive channels. Amplitudes for exclusive reactions and statistical ensembles in inclusive processes play complementary roles in bridging theoretical predictions with experimental data.

The \textit{S-matrix} encodes the transition between asymptotic states---free particle states in the distant past and future, before and after interactions take place. Formally, it is defined as the time-evolution operator in the limit of infinite time separation:
\begin{equation}
    \hat{S} = \lim_{t \to \infty} \hat{U}(t, -t)\ ,
\end{equation}
where $\hat{U}(t, t_0)$ is the time-evolution operator, defined by
\begin{equation}
    \hat{U}(t, t_0) |\psi(t_0)\rangle = |\psi(t)\rangle \ , \quad \text{with} \quad
    \frac{d}{dt} \hat{U}(t, t_0) = \frac{1}{i\hbar} \hat{H}_I \hat{U}(t, t_0)\,.
\end{equation}
Here, $\hat{H}_I$ is the interaction Hamiltonian, and $|\psi(t)\rangle$ is the state vector at time $t$. A hermitian Hamiltonian $\hat{H}_I$ ensures that the time-evolution operator $\hat{U}(t, t')$ is unitary, which in turn guarantees the unitarity of the $S$-matrix. This is the mathematical expression of probability conservation.

The $S$-matrix connects initial and final states via
\begin{equation} \label{eq:s-matrix}
    \langle \fin | \hat{S} | \ini \rangle = \delta_{\finini} + i(2\pi)^4 \delta^4(p_\fin - p_\ini) \Mcal_{\finini}\,,
\end{equation}
where $\delta_{f,i}$ represents the transition
for non-interacting particles (identical initial and final states),
and $\Mcal_{\finini}$ is the \textit{Lorentz-invariant scattering amplitude}.
The delta distribution $\delta^4(p_\fin - p_\ini)$ enforces the conservation of the total energy and momentum in the reaction with total four-momenta $p_\ini$ and $p_\fin$
for the initial and final states, respectively.
Scattering processes are formulated in terms of asymptotic momentum eigenstates (plane waves), effectively neglecting the spatial localization of wave packets~\cite{Peskin:1995ev}. This approximation avoids complications from wave-packet spreading while preserving essential features such as momentum transfer and angular dependence. The separation of the identity in the expression for the $S$-matrix is rigorously derived using the \textit{Lehmann-Symanzik-Zimmermann (LSZ) reduction formula}, which relates time-ordered Green's functions of interacting fields to $S$-matrix elements by amputating external propagators and taking \textit{on-shell} residues~\cite{Peskin:1995ev}.

Scattering amplitudes are not measurable directly. Experimentally accessible quantities are cross sections and decay rates.
For a scattering process of two particles, $A$ and $B$ with spins $j_A$ and $j_B$, respectively, into an $n$-body final state, the cross section is given by the squared magnitude of the amplitude integrated and summed over final state phase-space and spin variables and normalized by a flux factor.
It is computed as
\begin{equation} \label{eq:cross-section}
    \sigma = \frac{1}{J} \int \diff\Phi_n \,
    \frac{1}{(2j_A+1)(2j_B+1)}
    \sum_{\text{helicities}} |\Mcal_{\finini}|^2 \, ,
\end{equation}
where $\diff\Phi_n$ denotes the \textit{Lorentz invariant phase-space volume},
and the flux factor $J$ accounts for density and relative motion of the incoming particles,
\begin{equation}
    J = 4E_1 E_2 |v_A-v_B| = 2\lambda^{1/2}(s,m_A^2,m_B^2)\,.
\end{equation}
Here, $\lambda(x,y,z) = x^2+y^2+z^2-2xy-2yz-2xz$ is the K\"all\'en function~\cite{Kaldamae:2014fua},
and $v_A$, $v_B$ are the velocities of the incoming particles.
The phase-space integral sums over all possible momentum configurations
of the final-state particles that conserve energy and momentum~\footnote{Note that different conventions can be found in the literature. In particular, the factor $(2\pi)^4$ here appears as part of the phase space, while in other sources it appears as a prefactor.},
\begin{equation} \label{eq:phase-space}
    \diff \Phi_n = \prod_{i=1}^n \frac{\diff^3 k_i}{(2\pi)^3 2E_i} (2\pi)^4 \delta^4\left(\sum_{i=1}^n k_i - p_{\ini}\right) \,,
\end{equation}
where $p_{\ini}$ is the total momentum of the initial state, and $k_i$ are the momenta of the final-state particles.

For a narrow unstable particle in the initial state $i$, the partial decay width to the final state $f$ reads as
\begin{equation} \label{eq:decay_rate}
    \Gamma = \frac{1}{2m_0} \int \diff\Phi_n \, \frac{1}{2j_0+1}
    \sum_{\text{helicities}}|\Mcal_{\finini}|^2 \,,
\end{equation}
where $m_0$ and $j_0$ are the mass and spin of the decaying particle.
The same matrix-element squared governs the differential decay rate,
\begin{equation}
    \frac{\diff N_\mathrm{f}}{\diff \Phi_n} = Q \sum_{\text{helicities}}|\Mcal_{\finini}|^2 \,,
\end{equation}
where $\diff N_\mathrm{f}/\diff \Phi_n$ is the observed distribution of the final state kinematics and $Q$ is a constant that encodes the normalization and the production rate of the decaying particle.
For broad resonances, the value of $\Gamma$ might vary significantly over the mass of the system $m$ in the range \mbox{$m_0-\Gamma/2 < m < m_0+\Gamma/2$}. The width as a function of energy in Eq.~\eqref{eq:decay_rate} is commonly used to parameterize the imaginary part of the self-energy of the particle propagator---a connection provided by unitarity as detailed below.
The \textit{dimension} of the matrix element depends on the number of particles involved.
Cross sections have dimensions of an area, \textit{i.e.}, $m^{-2}$, while the partial widths have dimensions of inverse time, \textit{i.e.}, $m$. The dimension of the phase space is $m^{2n-4}$.
Hence, the amplitude has a dimension of $4-N$, where $N$ is the number of particles involved: $N = n+2$ for scattering and $N = n+1$ for decay.

The \textit{size of the amplitude} is driven by the underlying dynamics of the interaction and
can vary significantly depending on the kinematic variables.
Nonetheless, one can obtain a rough estimate using typical cross section values.
For strong interactions, the total proton-proton scattering
cross section at a center-of-mass energy of $100\,\text{GeV}$ is about $50\,\text{mb}$.
For electromagnetic interactions, $e^+ e^-$ scattering to hadrons gives an order of magnitude of about tens of $\text{nb}$.
For weak interactions, the cross section is much smaller, on the order of a few $\text{fb}$ for neutrino scattering~\cite{ParticleDataGroup:2024cfk}.

A recursive expression for constructing the phase-space integral for an $n$-particle
system reads as~\cite{Byckling:1971vca}:
\begin{equation}
    \diff\Phi_n (m_0; m_1\dots m_n) = \int \frac{\diff m^2}{2\pi}\,\diff\Phi_{n-1}(m; m_1\dots m_{n-1}) \,\diff\Phi_2(m_0; m, m_n)\,.
\end{equation}
The two-body phase space volume is given by
\begin{equation} \label{eq:rho.def}
    \diff \Phi_2(m; m_A, m_B) = \frac{1}{8\pi}\frac{2|\,\vec q\,|}{m} \frac{\diff \Omega}{4\pi}=2\rho(m^2) \quad \mbox{with}
    \quad   |\,\vec q\,| = \frac{\lambda^{1/2}(m^2,m_A^2,m_B^2)}{2m} \,.
\end{equation}
Here, the total energy is denoted by $m$ and the masses of the two particles by $m_A$ and $m_B$.
The abbreviation for the two-body phase space,
$\rho(m^2)$, is introduced here for later convenience.
Numerical sampling of the phase space is a common task in particle physics.
Multiple algorithms are based on the recursive decomposition of the phase space~\cite{Kleiss:1985gy,Platzer:2013esa}. While angles can be generated uniformly in a multi-dimensional square space, the physical domain for mass variables is a simplex.

It is instructive to count the variables that determine the dynamics of scattering and decay processes starting from the phase-space degrees of freedom.
Upon resolving the delta-function constraints in Eq.~\eqref{eq:phase-space}, one is left with $[\Phi_n] = 3n-4$ integration variables.
\begin{figure}
    \centering
    \includegraphics[scale=0.8]{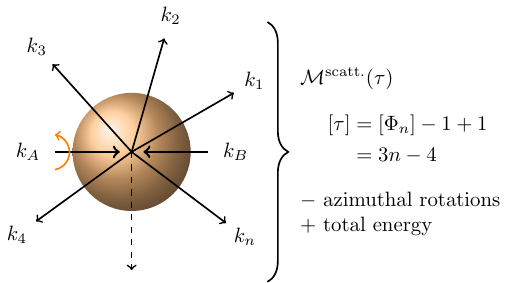}
    \includegraphics[scale=0.8]{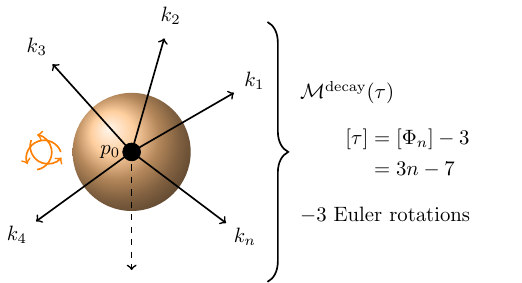}
    \caption{A sketch of momenta distribution for $n$-body final state in a scattering (left) and decay (right) kinematics.}
    \label{fig:phase-space-sketch}
\end{figure}
In scattering kinematics, once the initial momenta are fixed, there remains a freedom to perform an azimuthal rotation of the final-state momenta around the beam axis in the center-of-momentum frame, as illustrated in Fig.~\ref{fig:phase-space-sketch}. No observable in unpolarized process can depend on this rotation. The center-of-mass energy, in contrast, gives an additional variable to consider, since it's not counted in the degrees of the phase-space volume. This results in $N_\text{scatt.} = 3n - 4$ for the number of relevant kinematic variables.
A well-known consequence of this counting is that two-to-two scattering processes depend on two variables: the scattering angle and the center-of-mass energy, or equivalently, a pair of Mandelstam variables~\cite{Mandelstam:1958xc} introduced in the next section.

For a decay process, shown in the right panel of Fig.~\ref{fig:phase-space-sketch}, even more degrees of freedom are unphysical.
For unpolarized decay, probability density cannot depend on the overall orientation of the decay frame, specified by three Euler angles.
Consequently, the process depends on $N_\text{decay.} = 3n - 7$ variables.
This is consistent with representations like the Dalitz plot~\cite{Dalitz:1953cp} for three-body decays, where the full kinematic dependence is captured by two invariant masses of particle pairs.
This counting aligns with the general rule for the number of physical degrees of freedom in a process involving $N$ particles $N_\text{d.o.f.} = 3N - 10$, where 10 accounts for four-momentum conservation (4), three rotations (3), and three Lorentz boosts (3), which the amplitude cannot depend on.

The spin configuration of the amplitude can be tracked with the spin projection indices. \textit{Canonical quantization} of the spin is done with respect to a fixed axis, typically the $z$ axis.
\textit{Helicity} is the spin projection of a particle momentum onto its direction of motion.
For a decay process, the amplitude in the CM frame depends on helicity values of the final-state particles,
$\lambda_i$, and $m_0$, a canonical spin-projection value of the decay particle,
\begin{align} \label{eq:helicity}
    \text{decay}: \qquad \Mcal_{\finini}(\tau)       & = \Mcal_{\lambda_1,\ldots,\lambda_n; m_0}(\tau)\,,                 \\
    \text{scattering}:  \qquad \Mcal_{\finini}(\tau) & = \Mcal_{\lambda_1,\ldots,\lambda_n; \lambda_A,\lambda_B}(\tau)\,.
\end{align}
where $\tau$ denotes a set of kinematic variables.

When the initial state is unpolarized, it is natural to introduce an aligned kinematics by fixing the coordinate system using the momenta of the involved particles.
In scattering, the standard choice is to align the beam momentum with the $z$-axis, and to define the $xz$ plane by the momentum of one of the final-state particles~\cite{Gottfried:1964nx}. In three-body decays, all final-state particles lie in a single decay plane in the center-of-momentum frame.
The aligned amplitude describes the process in this specific frame,
involving the minimal number of degrees of freedom: $3n - 4$ for scattering and $3n - 7$ for decay.

The aligned amplitude is related to the amplitude in an arbitrary frame by a rotation of the coordinate system.
The general properties of the Lorentz group dictate how the amplitude depends on orientation angles,
independent of the underlying interaction dynamics~\cite{JPAC:2019ufm}.
For a system with fixed total spin $j$---for example, the decay of an unstable particle---the amplitude in an arbitrary frame is given by
\begin{equation} \label{resonances:eq:rotation}
    \Mcal_{\lambda_1,\ldots,\lambda_n; m}^{j}(\tau, \alpha,\beta,\gamma) =
    \sum_{m'} D^{j*}_{m,m'}(\alpha,\beta,\gamma)
    \Mcal^{j, (\text{aligned})}_{\lambda_1,\ldots,\lambda_n; m'}(\tau) \,,
\end{equation}
where $\alpha$, $\beta$, and $\gamma$ are the Euler angles defining the orientation of the decay plane.
The functions $D^{j}_{m,m'}(\alpha,\beta,\gamma)$ are Wigner $D$-matrices.
The rotation appears conjugated because the active rotation $R_z(\alpha),R_y(\beta),R_z(\gamma)$ is applied to the final-state momenta in the aligned frame to recover the general frame.
The spin projection $m'$ refers to the canonical quantization along the $z$-axis in the aligned configuration.

It is useful to introduce and discuss separately the \textit{production amplitude}, denoted $\Acal_{\finini}$. In contrast to the \textit{scattering amplitude} $\Mcal_{\finini}$, which describes transitions between states that are both part of the strongly interacting (not necessarily all of the strong-interaction) channel space, the production amplitude $\Acal_{\finini}$ describes transitions into hadronic channels when the initial state or effective source lies outside that sector. Decays, already discussed above, furnish one class of examples: the decaying particle acts as the source feeding the hadronic final state.
As a particular example, in three-body decays of $J/\psi \to \gamma\pi\pi$, one can isolate a two-body subsystem, $\pi\pi$, and treat its dynamics as a production amplitude, with the photon and the $J/\psi$ forming an effective source that couples only weakly to the strongly interacting $\pi\pi$ channel.
Further examples include an electromagnetic current ($e^+e^- \to \text{hadrons}$), a weak decay vertex, or any hadronic configuration with negligible feedback on the strongly coupled sector. Such a source can be viewed as an additional row and column of the full $S$-matrix, but its effect on the hadronic submatrix is negligible. It is therefore advantageous to restrict attention to the strongly coupled channels and collect the source transitions into a production amplitude vector $\Acal_{\finini}$, with one component per hadronic channel.

%% file: sections/partial-waves.tex
For scattering kinematics, the total spin of the system is not fixed.
The partial-wave expansion is a common technique used in that case to break down the full amplitude into components
with well-defined total spin $j$.
A scattering amplitude in the rest frame can be written as an infinite sum of partial waves,
\begin{equation} \label{eq:partial-wave-expansion}
    \Mcal_{\lambda_1, \ldots, \lambda_n; \lambda_a, \lambda_b} = \sum_{j=0}^{\infty} \Mcal_{\lambda_1, \ldots, \lambda_n; m}^{j} \,,
\end{equation}
where $m = \lambda_a - \lambda_b$ is the difference of helicities of the incoming particles in the CM frame.
Each term in the sum, $\Mcal_{\lambda_1, \ldots, \lambda_n; m}^{j}$, has a fixed total spin $j$, and therefore
a known angular dependence given by Eq.~\eqref{resonances:eq:rotation}.
The partial-wave expansion is of great practical importance for studying hadronic resonances,
since an intermediate resonance has a well-defined spin and, therefore, contributes to a single partial wave only.
It should be noted that the expansion provided in Eq.~\eqref{eq:partial-wave-expansion} can be applied to any subsystem of particles.
Such an expansion applied iteratively to multi-body reactions maps a general reaction to a cascade of intermediate states with a fixed spin---a procedure at the core of many partial-wave analyses~\cite{Hansen:1973gb,Herndon:1973yn}.

\begin{figure}
    \centering
    \raisebox{-0.5\height}{\includegraphics[scale=0.8]{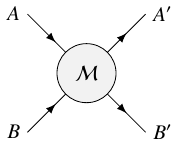}}\hspace{1cm}
    \raisebox{-0.5\height}{\includegraphics[scale=0.8]{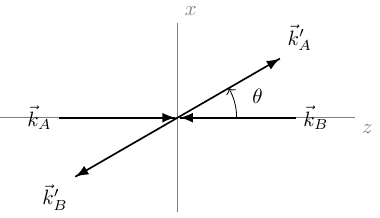}}
    \caption{Left: A diagram of the two-to-two scattering process $AB\to A'B'$.
        Right: kinematics of the scattering process in the center-of-mass frame, where the incoming particles $A$ and $B$ are moving along the $z$~axis, and the outgoing particles $A'$ and $B'$ are emitted at an angle $\theta$ with respect to the $z$~axis.}
    \label{fig:scattering}
\end{figure}

Let us now focus on two-to-two scattering processes, $AB\to A'B'$, to provide a more detailed discussion of the partial-wave expansion
that is of interest for the forthcoming discussion.
Here and below, the particles of a two-to-two process are labeled $A,B\to A',B'$, while $1,2,\ldots,n$ denote the particles of an $n$-body final state as in Fig.~\ref{fig:phase-space-sketch}.
Figure~\ref{fig:scattering} shows the \textit{scattering diagram} of such a process and illustrates the kinematics in the center of momentum frame.
The \textit{Mandelstam variables}
are defined via
\begin{equation}
    s = (k_A+k_B)^2, \ t=(k_A-k_{A'})^2, \ u=(k_A-k_{B'})^2\ .
\end{equation}
The variables $s$, $t$, and $u$ are not independent, as the relation
\begin{equation}
    s+t+u = m_A^2+m_B^2+m_{A'}^2+m_{B'}^2
\end{equation}
holds, where the $m_i$ with $i \in \{A,A',B,B'\}$ denote the masses of the particles involved.
As a result, the reaction amplitude for $2\to 2$ scattering can be expressed as a function of two variables,
$\Mcal(s,t)$.
The variable $t$ is related to the scattering angle $\theta$ in the center-of-mass frame by
\begin{align} \label{eq:t.cos}
    t(s, \cos\theta) & = m_A^2 + m_{A'}^2 -
    \frac{(s+m_A^2-m_B^2)(s+m_{A'}^2-m_{B'}^2)}{2s} +
    \frac{\lambda^{1/2}(s,m_A^2,m_B^2)\lambda^{1/2}(s,m_{A'}^2,m_{B'}^2)}{2s}\cos\theta\,.
\end{align}
In the absence of spins of the involved particles, the partial-wave expansion from Eq.~\eqref{eq:partial-wave-expansion} and Eq.~\eqref{resonances:eq:rotation} simplifies to
\begin{align}
    \Mcal(s,t) & =\sum_{j=0}^\infty (2j+1) \Mcal^j(s)P_{j}(\cos \theta) \,.
    \label{eq:pwdecomp}
\end{align}
Here, an additional numerical factor $(2j+1)$ is included for convenience, leading to
\begin{align}
    \int_{-1}^1 \frac{\diff \cos\theta}{2} \left|\Mcal(s,t)\right|^2 = \sum_{j=0}^\infty (2j+1)\left|\Mcal^j(s)\right|^2 \,.
\end{align}
The partial-wave amplitude $\Mcal^j$ is a function of the variable $s$ only. This leads to significant simplification of the analysis.
In practice, the expansion in Eq.~\eqref{eq:pwdecomp} is truncated and requires consideration of a sufficiently large number of partial waves.

%% file: sections/analyticity.tex
Causality is the principle that effects cannot precede their causes.
In quantum scattering theory, causality imposes powerful constraints on the scattering matrix.
In relativistic quantum field theory (QFT), \emph{microcausality}---the vanishing of commutators of field operators at spacelike separations---encodes the principle of locality~\cite{Bogolyubov:1975bv,Lehmann:1954rq}.
A striking consequence is that scattering amplitudes become \textbf{analytic functions} of complex kinematic variables,
such as energy or momentum invariants, in certain domains~\cite{Eden:1966dnq, Weinberg:1995mt}.
In particular, causal and local interactions imply that the S-matrix elements are free of singularities in parts of the complex plane, except for those required by physical thresholds and resonances or bound/virtual states.

\begin{figure}[h]
    \centering
    \raisebox{-0.5\height}{\includegraphics[width=0.44\textwidth]{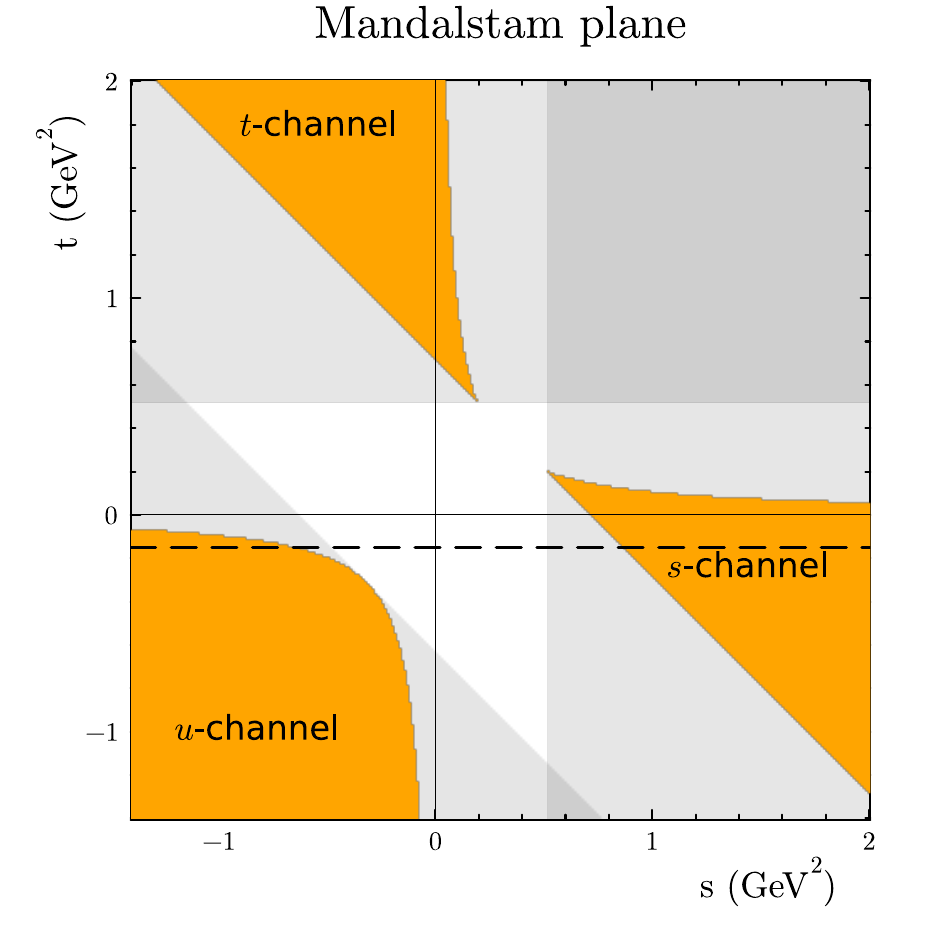}}
    \quad
    \raisebox{-0.5\height}{\includegraphics[width=0.44\textwidth]{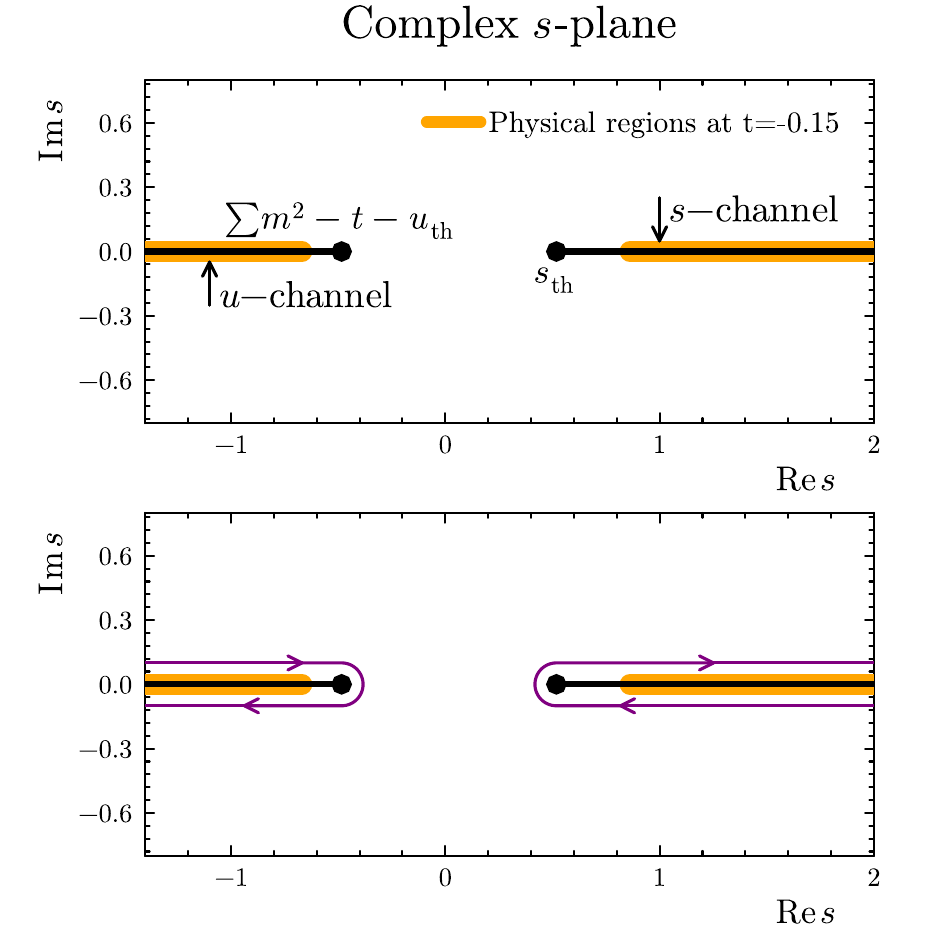}}
    \caption{\textbf{Left}: Mandelstam plane for $2\to 2$ scattering, using $\pi \eta$ scattering kinematics as an example. The amplitude is a real function in the white region below all physical thresholds. It acquires an imaginary part in the region shown in gray.
    The shaded region shows the region where the amplitude corresponds to a physical process.
    The dashed line gives an example for some fixed $t$ kinematics.
    \textbf{Right}: complex $s$-plane. Black solid markers indicate the branch points associated to the scattering thresholds,
    and the black lines give the branch cuts. Shaded regions indicate kinematic domains of the physical process for the
    value of $t$ shown in the panel.
    In this example, $m_A=m_{B'}=m_\eta$, $m_B=m_{A'}=m_\pi$. The thresholds for the $s$, $t$ and $u$ channels are $s_\text{th}=t_\text{th}=(m_\eta+m_\pi)^2$, and $u_\text{th}=4m_\eta^2$, respectively.
    The upper panel indicates how the physical regions with real Mandelstam variables for the $s$ and $u$ channels are approached from the complex plane,
    the lower one the integration contour used for the dispersion relations discussed in Eqs.~\eqref{eq:dispersion_relation}.
    }
    \label{fig:mandelstam_plane}
\end{figure}

This connection between microcausality and analyticity is realized concretely through the LSZ formalism, which relates S-matrix elements to time-ordered correlation functions of local fields. Because local field commutators $[\phi(x),\phi(y)]$ vanish at spacelike separation, the associated Green's functions possess analyticity properties in appropriate complex energy domains~\cite{Weinberg:1995mt,Bogolyubov:1975bv}. In free-field theory, this structure can be demonstrated explicitly. In interacting theories, the LSZ reduction formula expresses scattering amplitudes as on-shell limits of these same Green's functions, so that their analytic structure is inherited by the S-matrix. Although this causality-analyticity link underlies the standard framework of relativistic QFT, a fully rigorous nonperturbative proof for realistic interacting theories in 3+1 dimensions is still lacking. Much of our confidence stems from successful applications in simpler models, the fact that these properties are verified at finite order in perturbation theory, and compelling physical arguments rather than formal mathematical derivations.
The analytic continuation of scattering amplitudes into the complex plane introduces significant mathematical and conceptual challenges, particularly regarding the singularities of multivariate functions. The $2\to 2$ case, with its relatively simple analytic properties, provides a clear and instructive example for our discussion.
For this process, the amplitude is a function of two Mandelstam invariants, $s$ and $t$, as defined in the previous section. As illustrated in the left panel of Figure~\ref{fig:mandelstam_plane}, these variables define the Mandelstam plane, where different regions correspond to different physical processes. For stable particles one finds three scattering regions, associated with the $s$, $t$, and $u$ channels. If one particle is heavy enough to decay into the other three, a fourth physical region, the decay region, opens as well. Analyticity allows us to extend the amplitude into the complex plane, treating $s$ and $t$ as complex variables. This complex function exhibits branch cuts along the real axis, corresponding to physical thresholds associated with particle production.

In both variables, different physical regions manifest as singularities. The amplitude, originally defined for real $s$ and $t$, is related to its analytically continued counterpart through specific prescriptions for approaching the real axis. As shown in the right panel of Figure~\ref{fig:mandelstam_plane}, since the branch cut extends along the real axis, the values of the amplitude differ depending on whether the axis is approached from above or below. The physical scattering amplitude corresponds to one of these limits, which is captured by the $+\iepsilon$ prescription. The upper plot in the right panel specifically demonstrates how the physical regions with real Mandelstam variables for the $s$ and $u$ channels are approached from the complex plane.

Analytic properties of the amplitude in the complex plane provide powerful tools for understanding scattering processes. Extending the amplitude to the complex plane offers two key advantages. First, analytic functions are tightly constrained by their singularity structures, enabling systematic techniques from complex analysis. Second, placing the amplitude in a multivariable complex domain naturally incorporates \textit{crossing symmetry}: processes with different initial or final states can be viewed as analytic continuations of one another.

Dispersion relations provide a direct way to link different kinematic regimes of the amplitude, including its real and imaginary parts. A straightforward consequence of Cauchy's theorem is that an analytic function can be written as a contour integral around a closed path in the complex plane. By extending this contour to infinity and applying suitable conditions for the UV behavior of the integrand, one obtains a method for evaluating the amplitude at any point in the complex plane, given its discontinuities along the real axis. This integration procedure is illustrated in the lower plot of the right panel in Figure~\ref{fig:mandelstam_plane}, which shows the integration contours used in dispersion relations.
Concretely, for some fixed value of $t$, the relation is
\begin{equation} \label{eq:dispersion_relation}
    \mathcal{M}(s, t) = \frac{1}{2\pi i} \int_{s_\text{th}}^{\infty} \frac{\Delta_s\mathcal{M}(s', t)}{s'-s} \diff s' + \frac{1}{2\pi i} \int_{s(u_\text{th},t)}^{-\infty}\frac{\Delta_u\mathcal{M}(s', t)}{s'-s} \diff s'\,,
\end{equation}
where $s(u_\text{th},t) = m_A^2+m_B^2+m_{A'}^2+m_{B'}^2 - t - u_\text{th}$ determines the $u$-channel physical threshold in the $s$ plane for the given $t$.
The discontinuities are
\begin{align} \label{eq:dispersion.relation.two}
    \Delta_s\mathcal{M}(s', t) & = \mathcal{M}(s'+\iepsilon, t) - \mathcal{M}(s'-\iepsilon, t)\,, \\
    \Delta_u\mathcal{M}(s', t) & = \mathcal{M}(s'-\iepsilon, t) - \mathcal{M}(s'+\iepsilon, t)\,.
\end{align}
The opposite sign in the $u$-channel discontinuity arises because its physical region $u+\iepsilon$ lies in the lower half-plane of the $s$ variable.

An immediate consequence of analyticity is that any feature on the real axis, such as peaks, dips, or cusps, originates from singularities in the complex plane.
Unless the \textit{left-hand cut} and \textit{right-hand cut} overlap~\cite{Eden:1966dnq,Mandelstam:1958xc}, there is a segment of the real axis where the amplitude is real.
Then, the analytic continuation of the amplitude to the complex plane follows from the \textit{Schwarz reflection principle}:
\begin{equation} \label{eq:schwarz_reflection}
    \mathcal{M}(s_x^*, t) = \mathcal{M}(s_x, t)^*\,.
\end{equation}
This implies that the discontinuity across the cut equals twice the imaginary part of the function above the cut.

An example of an amplitude that has only a single right-hand branch cut is the pion vector form factor measured in $e^+e^-\to \pi^+\pi^-$.
Then, the dispersion relation from Eq.~\eqref{eq:dispersion.relation.two} reduces to
\begin{equation} \label{eq:dispersion_relation_simple}
    \mathrm{Re}\,\mathcal{F}(s) = \frac{1}{\pi}\,\mathcal{P}\!\int_{s_\text{th}}^{\infty} \frac{\mathrm{Im}\, \mathcal{F}(s')}{s' - s}\,\diff s'\,,
\end{equation}
where $\mathcal{P}$ denotes the Cauchy principal value. This result is analogous to the \textit{Kramers--Kronig relations}~\cite{Kramers:1927,Kronig:1926}.

Both Eqs.~\eqref{eq:dispersion_relation} and \eqref{eq:dispersion_relation_simple} may require subtractions to ensure convergence of the integral. If the function saturates to a constant at high energies, a single subtraction suffices, which corresponds to applying the dispersion relation to $(F(s) - F(s_\text{sub})) / (s - s_\text{sub})$, where $s_\text{sub}$ is the subtraction point.
For multivariable complex functions, analytic properties are significantly more intricate than in the single-variable case. Even in the $2\to 2$ scattering example, once the variable $t$ moves into its physical region, the left and right branch cuts begin to overlap, removing the real analytic segment. The dispersion contour must then be deformed into the complex plane; for partial-wave amplitudes with unequal masses the cut structure is even richer~\cite{Kubis:2025zji}.

Rather than attempting to derive analyticity and crossing symmetry directly from a fundamental theory, modern approaches take these properties as guiding principles. Dispersion relations have led to the derivation of \textit{positivity bounds} in effective field theory~\cite{Adams:2006sv}, while numerical \textit{$S$-matrix bootstrap} methods use analyticity, crossing symmetry, and unitarity to constrain possible scattering amplitudes as described in~\cite{Paulos:2016fap}. Even in the absence of a complete proof linking causality to analyticity in general quantum field theories, these methods demonstrate the enduring power of analyticity in shaping our understanding of particle interactions.

%% file: sections/unitarity.tex
\begin{figure}[t]
    \centering
    \includegraphics[scale=0.8]{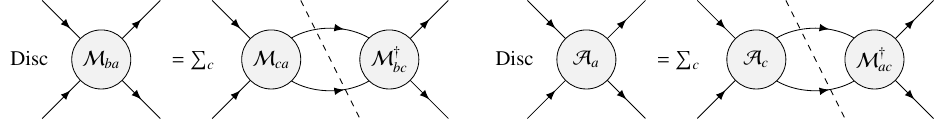}
    \caption{
        Graphical illustration of the discontinuity equations for
        the scattering and the production amplitude, respectively.
        The dashed line indicates that the intermediate state is to be
        put on-shell to find the discontinuity.
        The figure is adapted from Ref.~\cite{ParticleDataGroup:2024cfk}.
    }
    \label{fig:discs}
\end{figure}

The unitarity of the $S$-matrix implies a condition on reaction amplitudes,
\begin{equation}
    \Mcal_{ba}(s, t)-\Mcal_{ba}^\dagger(s, t) =
    i \,\sum_c \int \diff \Phi_c \Mcal_{bc}^\dagger(s, t')\Mcal_{ca}(s, t'') \ ,
    \label{eq:unitarity}
\end{equation}
where $\Mcal_{ba}^\dagger(s, t)$ is defined from the expectation for the $S^\dagger$ operator analogous to Eq.~\eqref{eq:s-matrix}. Indices $a$, $b$, and $c$ label channels that are coupled to each other by scattering processes.
The intermediate state $c$ that appears under the integral is a set of on-shell multi-particle states that can connect $a$ to $b$: the Lorentz-invariant phase space of Eq.~\eqref{eq:phase-space} puts every particle of channel $c$ on its mass shell.
Unitarity is a constraint at a given energy $\sqrt{s}$. Accordingly, the sum in Eq.~\eqref{eq:unitarity} runs only over \textit{open} channels, those whose production threshold lies below the energy of the scattered system~\cite{ParticleDataGroup:2024cfk,Oller:2020guq}.
One can enforce this restriction explicitly by inserting a Heaviside function $\Theta(s-s_{\mathrm{th};c})$ in Eq.~\eqref{eq:unitarity}.
The energy variable $s$ is the same for all channels, while the variables $t$, $t'$ and $t''$ are defined for each amplitude as the transferred momentum between its respective initial and final states.
The integration over configurations of the intermediate state $c$ affects the $t'$ and $t''$ variables.
As long as time reversal invariance holds, $\Mcal$ is symmetric.
The relation is illustrated in the left part of Fig.~\ref{fig:discs}.
The known form of the \textit{optical theorem} is obtained
by considering the elastic amplitude, $\Mcal_{aa}(s, t)$.
For forward scattering, variables $t'$ and $t''$ that appear under the integral become the same.
Using the definition of the cross-section in Eq.~\eqref{eq:cross-section}:
\begin{align} \label{eq:optical.elastic}
    \mathrm{Im}\,\Mcal_{aa}(s, 0) & =
    (2j_A+1)(2j_B+1) \cdot 2q_a \sqrt{s} \sum_c \sigma_{ca}(s)\,,
\end{align}
where we used indices $A$ and $B$ for particles of channel $a$.
Here, $q_a$ is the center-of-mass momentum from Eq.~\eqref{eq:rho.def}.


In the partial-wave basis following the partial-wave expansion in Eq.~\eqref{eq:pwdecomp},
the unitarity relation from Eq.~\eqref{eq:unitarity} takes an algebraic form,
\begin{align}\label{eq:unitarity.pw}
    \mathrm{Im}\,\Mcal_{ba}^j(s) & =
    \sum_c\Mcal_{bc}^{j*}(s)\,\rho_c(s)\,\Mcal_{ca}^j(s)\,,
\end{align}
where we use the short-hand notation $\rho_c(s)$, introduced in Eq.~\eqref{eq:rho.def},
for the phase-space factor of two-body channels.
As in Eq.~\eqref{eq:unitarity}, the sum includes only open channels.

The unitarity relation derived on the real axis extends naturally into the complex plane due to the analytic structure of the scattering amplitude. The discontinuity of the amplitude across the real axis in the complex $s$-plane is a direct consequence of unitarity. The branch cut reflects the onset of physical thresholds, and unitarity determines the discontinuity across it via Eq.~\eqref{eq:unitarity}. Specifically, the physical amplitude $\Mcal_{ba}^j(s)$ corresponds to the boundary value from above the cut, $\Mcal_{ba}^j(s + \iepsilon)$, while $\Mcal_{ab}^{j*}(s)$, appearing in the unitarity relation, is understood as $\Mcal_{ba}^j(s - \iepsilon)$. This identification follows from analyticity and CPT invariance and does not require time-reversal or crossing symmetry~\cite{resonances:Olive:1962xyz}. Thus, unitarity alone determines the discontinuity, and the structure of the cut is a direct and necessary consequence of unitarity.


The related discontinuity equation for production reactions is shown on the right of Figure~\ref{fig:discs}.
It may thus be written as:
\begin{equation}
    \Acal_a(s,t)-\Acal_a^\dagger(s,t)
    = i \,\sum_c \int \diff \Phi_c \Mcal^\dagger_{ac}(s,t')\,\Acal_c(s,t'') \,.
    \label{eq:discA}
\end{equation}
Equation~\eqref{eq:discA} shows the close link between scattering and production amplitudes.
It's also important to note that the poles that appear in a given subsystem of
a production reaction are necessarily identical to those of the scattering amplitude.
In the partial-wave basis, the unitarity relation is algebraic and linear in the production amplitude,
\begin{equation}
    \mathrm{Im}\,\Acal_a^j(s) = \sum_c \Mcal^{j*}_{ac}(s)\,\rho_c(s)\,\Acal_c^j(s) \,.
    \label{eq:discA.pw}
\end{equation}
Moreover, if the final state interaction is purely elastic such that there is no summation over the intermediate channel index $c$,
Eq.~\eqref{eq:discA.pw} provides a proof of the Watson theorem~\cite{Watson:1952ji,Watson:1954uc},
namely that for elastic interactions the phase of the production amplitude agrees with the phase of the scattering amplitude.
To see this, simply observe that the imaginary part of the production amplitude, given by the expression, $\Mcal^{j*}_{aa}(s)\,\rho_a(s)\,\Acal_a^j(s)$, is real.

When several pairwise subsystems of a three-body final state can rescatter, the production amplitudes in different two-body channels are themselves coupled by crossing. A systematic framework that implements two-body unitarity in all three subsystems simultaneously is provided by the Khuri--Treiman equations~\cite{Khuri:1960zz,Aitchison:1966lpz,Aitchison:2015jxa}; a pedagogical discussion is given in Ref.~\cite{Kubis:2025zji} in this Encyclopedia.

%% file: sections/continuation.tex
Scattering amplitudes are multivalued functions of energy, naturally defined on multiple sheets of complex plane referred to as \textit{Riemann sheets}.
However, from a practical standpoint, the analytic structure and analytic continuation can be more easily understood by considering complex functions $\MI(s)$, $\MII(s)$, $\MIII(s)$, each defined on a single Riemann sheet of the complex plane. For each complex value of the energy variable $s$, the amplitude either has a well-defined value or encounters a singularity, such as a pole, a branch point, or a cut.\footnote{The specific location of branch cuts often depends on the numerical implementation rather than purely mathematical properties. For instance, the mathematically equivalent functions $\sqrt{z^2 - 1}$ and $\sqrt{z - 1}\sqrt{z + 1}$ differ numerically due to their distinct branch cut placements.}
These related single-valued functions analytically continue one another across branch cuts.
\begin{figure}[t]
    \centering
    \includegraphics[width=0.92\linewidth]{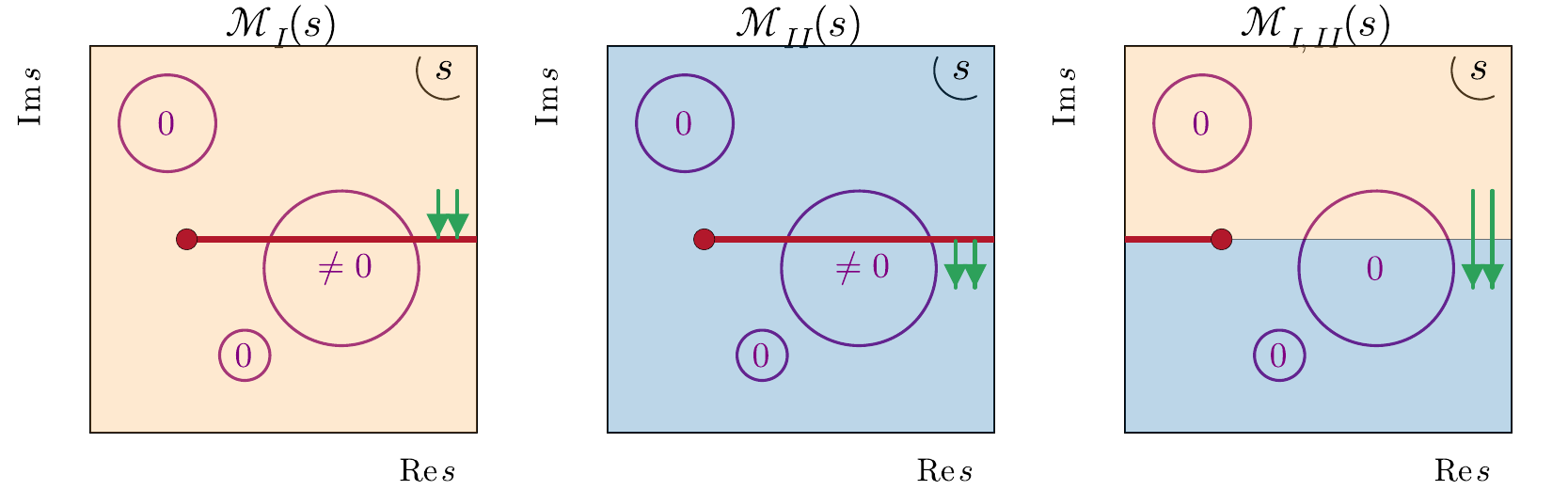}
    \caption{Demonstration of how the analytic domain of the function $\MI(s)$ is extended by a function $\MII(s)$.
        $\MIandII(s)$ is defined as $\MI(s)$ for $\mathrm{Im}\,s>0$, and as $\MII(s)$ for $\mathrm{Im}\,s<0$.
        Plots represent the complex $s$ plane. Domains of analyticity are shown for $\MI(s)$, $\MII(s)$, and $\MIandII(s)$ in the left, middle, and right panels, respectively. Black lines indicate the branch cuts, along which the functions are discontinuous.
        The black dots represent the branch points. Closed contour integrals vanish for analytic functions unless they enclose singularities.
        Purple circles give examples of the closed-contour integration paths, and annotations indicate whether the integral vanishes.}
    \label{fig:analytic_join}
\end{figure}
An illustrative example is shown in Fig.~\ref{fig:analytic_join}, where the function $\MI(s)$ has a cut starting from a threshold $s_{\mathrm{b.p.}}$ along the real axis, and $\MII(s)$ has a similar cut. We say $\MII(s)$ analytically continues $\MI(s)$ when the two complex planes, joined along the cut, form a continuous analytic manifold for a combined function $\MIandII(s)$, defined to be equal to $\MI(s)$ for complex $s$ with a positive imaginary part, and $\MII(s)$ for complex $s$ with a negative imaginary part.
The notation $\MI(s)$ is commonly used for the scattering amplitude on the \textit{physical Riemann sheet}. It is a single-valued analytic function defined on the complex plane with no singularities off the real axis.
The direction of open-channel branch cuts is dictated by analyticity: they extend to the right from the branch points.%
\footnote{The combined function $\MIandII(s)$ revealing both the first and second Riemann sheets is more than a mere mathematical construct. Such representations frequently arise in numerical implementations of phenomenological parametrizations, such as the Breit-Wigner amplitude, $\mathrm{BW}(s) = 1/(m^2 - s - i g^2 \rho(s))$.}

Each open channel corresponds to a branch cut, starting from a branch point located at the channel threshold. Two-body thresholds manifest as square-root branch points. Analytic continuation along a two-body unitarity cut is computed using the relation:
\begin{equation}\label{eq:disc.Mcal}
    \text{Disc}\,\Mcal^{-1}(s) = \Mcal^{-1}(s+\iepsilon) - \Mcal^{-1}(s-\iepsilon) = -2i\rho(s)\,,
\end{equation}
where $\rho(s)$, given by Eq.~\eqref{eq:rho.def}, satisfies $\rho(s+\iepsilon)=\rho(s-\iepsilon)$. It has two branch points: a threshold at $s_{\text{th}} = (m_A+m_B)^2$ and a pseudothreshold at $s_{\text{pth}} = (m_A-m_B)^2$.
Following Eq.~\eqref{eq:disc.Mcal}, one constructs the function,
\begin{equation}
    \MII^{-1}(s) = \MI^{-1}(s) - 2i\rho(s)\,.
\end{equation}
Indeed, the value of $\MII^{-1}(s)$ just below the cut on its complex plane matches the value of $\MI^{-1}(s)$ just above the cut:
\begin{equation}
    \MII^{-1}(s - \iepsilon) = \MI^{-1}(s - \iepsilon) - 2i\rho(s) = \MI^{-1}(s + \iepsilon)\,.
\end{equation}

\begin{figure}[t]
    \centering
    \includegraphics[width=0.33\linewidth]{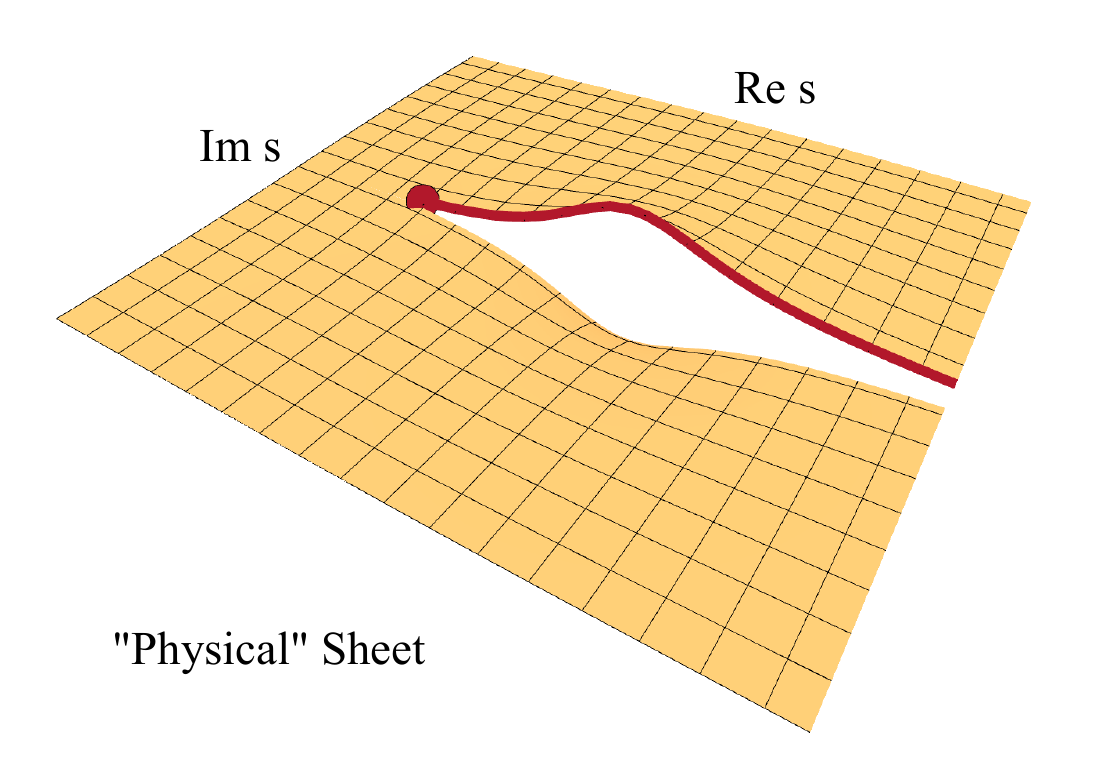}  
    \includegraphics[width=0.33\linewidth]{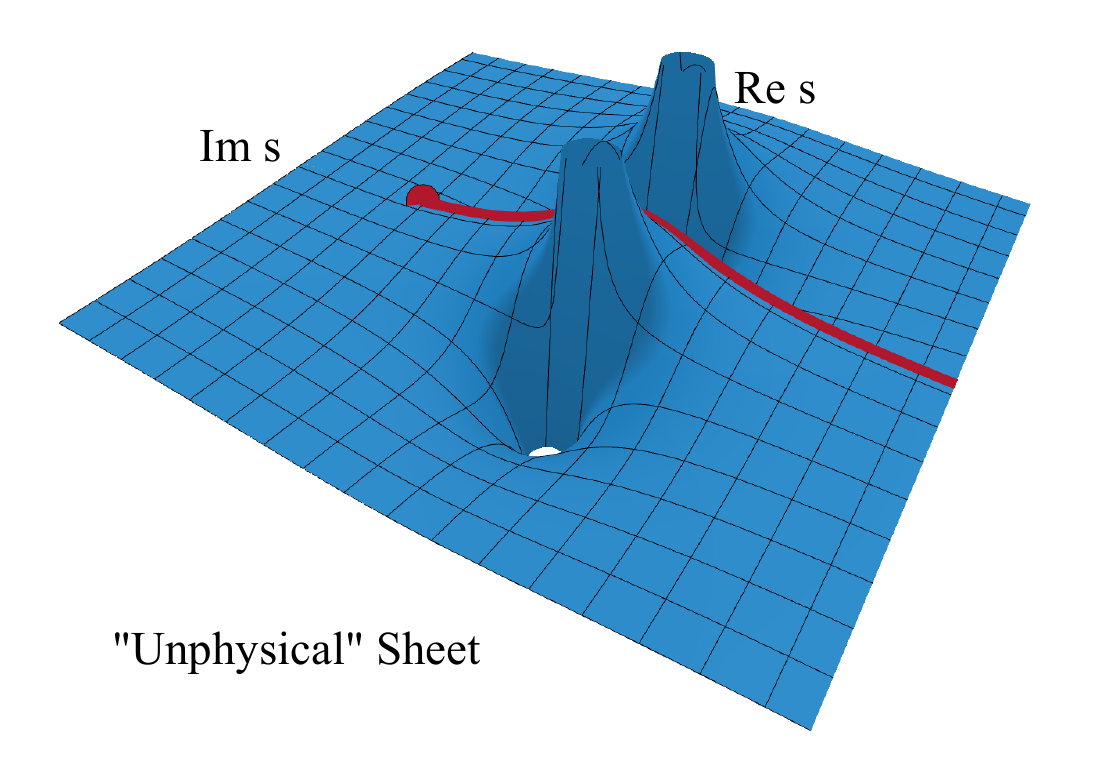} 
    \includegraphics[width=0.33\linewidth]{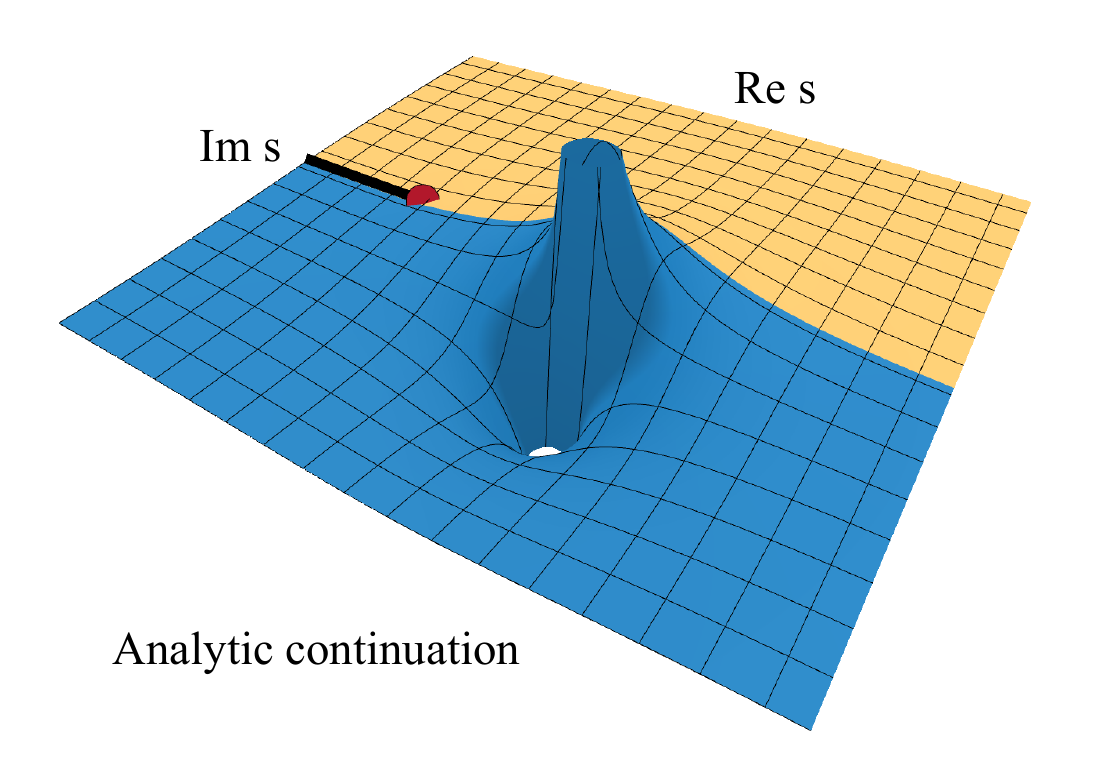}
    \caption{
        Illustration of the effect of a resonance in a one-channel amplitude. A resonance appears as a pair of poles on the second sheet (panel (b)) in the complex $s$-plane, which imprints
        the physical axis shown as the red line in panel (a), which shows the first or physical sheet. Panel (c) shows how the two sheets are connected.
    }
    \label{fig:sheets12}
\end{figure}

Most hadrons appear as resonances in scattering processes.
These resonances are characterized by poles of the scattering amplitudes.
The poles located on the real axis of the physical sheet below the lowest threshold are called \textit{bound states}.
Poles on unphysical sheets could be located on the real axis (then called \textit{virtual states}) or in the complex plane (referred to as resonances).
If there is only one relevant channel, the $s$-plane is split into two sheets.
The appearance of resonance poles on the unphysical sheet is then illustrated in Fig.~\ref{fig:sheets12}.
The singularity inside the complex plane on the unphysical sheet in blue induces a bump on the physical axis, which is part of the physical sheet shown in orange.
For the setting shown in Fig.~\ref{fig:sheets12} it should be clear that the second pole (the one in the upper half plane) plays at most a minor role close to the resonance peak; however, near threshold both poles are equally important.
The presence of the complex-conjugated poles is a consequence of the Schwarz reflection principle, and generalizes to higher sheets.

The location of the resonance pole is quoted to characterize the particle. The \textit{pole mass} and \textit{pole width} are defined as
\begin{equation}
    \sqrt{s_\res} = m - i\Gamma/2 \,,
\end{equation}
and are measured for many hadrons~\cite{ParticleDataGroup:2024cfk}.
The strengths of the amplitude at the pole for the different channels are determined by the residues, defined as
\begin{equation}
    \lim_{s\to s_\res}(s-s_\res)\Mcal_{ba} = - \Rcal_{ba} \, .
    \label{resonances:eq:resdef}
\end{equation}
Those can be conveniently calculated via an integration along a closed contour around the pole using the residue theorem:
\begin{equation}
    \Rcal_{ba} = -\frac{1}{2\pi i}\oint \diff s \, \left[\MII(s)\right]_{ba} \ .
\end{equation}
The residue matrix is subject to the factorization relation,
\begin{equation}
    \Rcal_{ab}^2 = \Rcal_{aa}\times \Rcal_{bb}\,.
    \label{resonances:eq:factorization}
\end{equation}
This relation, which emerges as a universal property from the unitarity of scattering processes~\cite{Gribov:2009zz},
allows one to define \textit{pole couplings} as,
\begin{equation}
    \tilde{g}_a \tilde{g}_b = \Rcal_{ab}\,.
    \label{resonances:eq:gpoledef}
\end{equation}
We note that care needs to be taken when computing the square root of the diagonal elements of the residue matrix and picking the sign consistent with Eq.~\eqref{resonances:eq:gpoledef}.
The pole couplings $\tilde{g}_a$ characterize the transition strengths of a given resonance to channel $a$, independently of how the particular resonance was produced. They are generally complex valued.

%% file: sections/theoretical.tex

The non-linear character of Eq.~\eqref{eq:unitarity.pw} shows that
no perturbative calculation can fulfill the unitarity requirements exactly.
Unitary amplitudes can be viewed as the solution of scattering equations that in operator form can be written as
\begin{equation}
    \Mcal =  \Vcal + \Vcal \Gcal \Vcal + \Vcal \Gcal \Vcal \Gcal\Vcal + \cdots = \Vcal + \Vcal \Gcal \Mcal \ ,
    \label{eq:BSE}
\end{equation}
where \Vcal is a scattering kernel and \Gcal is a Green's function operator.
In general, when evaluated in the plane-wave basis,
the kernel is a function of the ingoing and outgoing four-momenta.
\begin{equation}
    \langle A'B'| \Vcal | A,B\rangle = \Vcal_{A'B';AB}(s,t; k_{A}^2,k_{B}^2,k_{A'}^2,k_{B'}^2) \, (2\pi)^4 \delta^4(p_{A'B'}-p_{AB})\,.
    \label{eq:full.potential}
\end{equation}
The Green's function \Gcal is an operator acting on the space of particle states. For two-body scattering, it is the product of two single-particle propagators,
\begin{equation}\label{eq:Gdef}
    \Gcal_a(k_A,p) =
    \frac{i}{k_A^2-m_{A}^2+\iepsilon}\frac{i}{k_B^2-m_{B}^2+\iepsilon} \,.
\end{equation}
The index $a$ indicates the scattering channel; the Green's function is diagonal in the channel space.
The product of operators in Eq.~\eqref{eq:BSE} implies
integration over the intermediate momenta, $i\,\diff^4k_A / (2\pi)^4$.
Because of this integral, the potential that enters the scattering equation is probed not only on-shell ($m_i^2=k_i^2$ for $i=A,B,A',B'$) but also off-shell.
For relativistic two-body systems, Eq.~\eqref{eq:BSE} represents a
set of four-dimensional integral equations called
the \textit{Bethe-Salpeter equations}. The perturbative version of it reproduces
the well-known \textit{Born series}.
The equation~\eqref{eq:BSEsol} can be written as
\begin{equation}
    \Mcal = \left[1- \Vcal \Gcal\right]^{-1} \Vcal\,,
    \label{eq:BSEsol}
\end{equation}
where the inverse acts in continuous four-momentum space. Its evaluation therefore amounts to solving a system of coupled integral equations.
Nonetheless, Eq.~\eqref{eq:BSEsol} is instructive. Below, we use it to demonstrate that the solution of the Bethe-Salpeter equation satisfies two-body unitarity once the potential is real-valued, and the discontinuity induced by $\Gcal$ is given by $2i\rho_a(s)$. The demonstration is based on the on-shell approximation, but is true in general.

If the potential contains finite-range interactions mediated by,
e.g., the exchange of mesons,
it is not separable and the scattering amplitude inherits left-hand cuts.
Then, the four-dimensional version of Eq.~\eqref{eq:BSE}
is very difficult to solve numerically
(for more details see Section ``Hadron physics with functional methods'').
In Ref.~\cite{Lahiff:1999ur}, this is demonstrated for the pion-nucleon system.
To avoid these complications,
often either certain recipes are applied
to convert Eq.~\eqref{eq:BSE} into a three-dimensional integral equation~\cite{Gross:1993zj},
or the scattering amplitudes are directly constructed using time-ordered
perturbation theory, which automatically gives three-dimensional
integral equations that can be solved with standard tools,
see an example in Ref.~\cite{Ronchen:2012eg}.
The former brute-force three-dimensional reduction method discussed in Ref.~\cite{Gross:1993zj} renders
covariant equations that, however, suffer from unphysical singularities.
The latter method is free of those singularities,
but violates covariance, which, however, can be restored systematically~\cite{Zhang:2021hcl}.
Moreover, the irreducible contributions of successive two-meson exchanges that emerge in the time-ordered perturbation theory, as so-called stretched boxes, are numerically identical to the related crossed boxes that are not generated from
the four-dimensional scattering equation~\cite{Machleidt:2011zz,Chacko:2024cax}.
Note that the mentioned approximations can be controlled quantitatively
in tailor-made effective field theories as is demonstrated for few-nucleon systems in Ref.~\cite{Epelbaum:2008ga}.
In cases where only a small
energy range is relevant or the dynamics is dominated by
$s$-channel poles, the interaction can be treated as
separable allowing for a closed-form solution
of Eq.~\eqref{eq:BSE}---see, e.g. Ref.~\cite{Baru:2010ww,Guo:2016bjq}.
For systems without left-hand cuts, often the so-called on-shell approximation is employed,
where the kernel of Eq.~\eqref{eq:full.potential}
gets replaced by the on-shell potential by fixing
$k_i^2=m_i^2$---formally, the difference can be absorbed into short-range operators~\cite{Oller:1997ti}.%
\footnote{The mentioned on-shell approximation is often employed in the analysis of Lattice QCD data~\cite{Briceno:2017max,Mai:2022eur}.}
Then propagator and potential factorize, and the former can be replaced by a simple, universal one-loop integral,
\begin{equation} \label{eq:Sigma.def}
    \Chew_a(s) = i \int \frac{d^4k}{(2\pi)^4}\, \Gcal_a(k, p) = i \int \frac{d^4k}{(2\pi)^4}
    \frac{i}{k^2-m_A^2+\iepsilon}
    \frac{i}{(p-k)^2-m_B^2+\iepsilon}\,.
\end{equation}
This function plays a central role in phenomenological studies of hadron dynamics and largely determines the analytic structure of hadronic scattering amplitudes.
We may find the discontinuity of the two-particle Green's function by putting the particles
in the intermediate state of $\Gcal_a$ on their respective mass shells by employing the \textit{Cutkosky rules}~\cite{Cutkosky:1960sp},
\begin{equation}
    \frac{1}{k^2-m^2+\iepsilon} \to -2\pi i \,\delta(k^2-m^2)\theta(k_0)\,,
\end{equation}
for the two propagators appearing in Eq.~\eqref{eq:Sigma.def}.
With this we get\footnote{This is most easily demonstrated for $\vec p=0$ and $p^2=s$, but holds for all frames.}
\begin{equation}
    \mathrm{Disc}\,\Chew_a(s) = i\int \frac{d^4k}{(2\pi)^2} \, \delta(k^2-m_A^2) \, \delta((p-k_A)^2-m_B^2)
    =2i \rho_a(s)\,.
    \label{eq:discG}
\end{equation}
With a straightforward manipulation, one can show that the integral is equal to the two-particle phase space, introduced in Eq.~\eqref{eq:rho.def}.

The self-energies $\Chew_c$ can be written
as a once-subtracted dispersion integral over
the discontinuity of $\Chew$~\cite{Basdevant:1977ya},
\begin{eqnarray}\nonumber
    \Chew_a(s) &=& \frac{s-s_{\thr_a}}{\pi}\int_{s_{\thr_a}}^{\infty}\frac{
    \rho_a(s') }{(s'-s_{\thr_a})(s'-s-\iepsilon)} \,\diff s' \\
    &=& \frac{1}{16\pi^2}\left[\frac{2q_a}{\sqrt{s}}
    \log\frac{m_{A}^2+m_{B}^2 - s + 2\sqrt{s}q_a}{2 m_{A} m_{B}}  - \left(\frac{(m_{A} - m_{B})((m_A+m_B)^2-s)}{s(m_A+m_B)}\right) \log\frac{m_{A}}{m_{B}}
    \right]
    \,,
    \label{eq:Chew}
\end{eqnarray}
where $q_a$ is the break-up momentum defined in Eq.~\eqref{eq:rho.def}. The expression is defined everywhere in the complex plane except for the branch cut along the real axis starting at $s_{\thr_a}=(m_A+m_B)^2$.
This expression is the $S$-wave version of the famous Gounaris-Sakurai expression
introduced for vector mesons in Ref.~\cite{Gounaris:1968mw}.
In general, for partial waves larger than zero, centrifugal barrier factors (which lead to an additional momentum dependence proportional to $q_c^\ell$ at the pertinent vertices) and suppression factors typically of Blatt-Weisskopf type~\cite{Blatt:1952ije} need to be included in the dispersion integral to tame the high-energy behavior of the self-energy.
For a more detailed discussion, see the Resonance Review in Ref.~\cite{ParticleDataGroup:2024cfk}. An explicit form of the resulting self-energy for $\ell=1$ is provided in Ref.~\cite{Du:2025beb} that can be extended to higher partial waves straightforwardly.

To make contact between Eq.~\eqref{eq:BSE} and Eq.~\eqref{eq:unitarity.pw},
note that by construction $\Vcal$ is real-valued and does not contribute to the imaginary part of the scattering amplitude. With this,
we find
\begin{align} \nonumber
    \mathrm{Disc}\,\Mcal & = \Mcal-\Mcal^*= \Vcal \Chew \Mcal-(\Vcal \Chew \Mcal)^*
    =\Vcal \Chew \Mcal-\Vcal \Chew^* \Mcal+\Vcal \Chew^* \Mcal-\Vcal \Chew^* \Mcal^*                           \\
                         & =\Vcal \, \mathrm{Disc}(\Chew) \Mcal + \Vcal \, \Chew^* \, \mathrm{Disc}(\Mcal) \,,
\end{align}
where the equations are written for the on-shell scattering amplitude,
and the term product denotes matrix multiplication in the channel space.
Solving for $\mathrm{Disc}\,\Mcal$ and using Eq.~\eqref{eq:BSEsol} reproduces
Eq.~\eqref{eq:unitarity.pw}. The algebraic manipulations are straightforward in the one-channel case, but the relation also generalizes to coupled channels.
The scattering kernel must be two-body irreducible, which
means that it contains only those contributions that do not
feature the two-body discontinuity that is contained in $\Gcal$, see Eq.~\eqref{eq:discG}.

The expression analogous to Eq.~\eqref{eq:BSE} for production amplitudes reads
\begin{equation} \label{eq:BSE.production}
    \Acal = \Pcal + \Mcal \Gcal \Pcal \,,
\end{equation}
where $\Pcal$ denotes a reaction-specific source operator, which is free of right-hand cuts, \textit{i.e.} it is real-valued along the right-hand cut.
One can show that Eq.~\eqref{eq:BSE.production} satisfies the unitarity relation in Eq.~\eqref{eq:discA.pw} for general off-shell amplitudes; in practice, however, the on-shell approximation is employed, which removes the integral and replaces $\Gcal$ by the universal one-loop integral, Eq.~\eqref{eq:Sigma.def}.

Employing Eqs.~\eqref{eq:unitarity.pw} and \eqref{eq:discG}, we find
\begin{align} \nonumber
    \mathrm{Disc}\,\Acal & =  \Mcal \Chew \Pcal -  \Mcal^* \Chew^* \Pcal
    =  \Mcal \Chew \Pcal -  \Mcal^*   \Chew \Pcal +  \Mcal^* \Chew \Pcal -  \Mcal^* \Chew^* \Pcal          \\ \nonumber
                         & =  \mathrm{Disc}(\Mcal)\,\Chew\, \Pcal + \Mcal^*\,  \mathrm{Disc}(\Chew)\,\Pcal \\
                         & = 2i\Mcal^* \rho\,\Acal \ ,
\end{align}
in line with Eq.~\eqref{eq:discA}. This shows nicely how unitarity links the discontinuity of the production operator
to that of the scattering amplitude and how those features map onto dynamical equations.

Contrary to a perturbative treatment,
the fully resumed Bethe-Salpeter equation, Eq.~\eqref{eq:BSE},
bears the potential to generate poles in the scattering amplitude.
These poles appear whenever the equation
\begin{equation}
    \mathrm{det}\left[1- \Vcal \Chew\right]=0
    \label{eq:poles.I}
\end{equation}
has a solution.
Due to analyticity constraints, Eq.~\eqref{eq:poles.I} can only have solutions at real values of $s$ on the first sheet.
In addition, there might be
poles in the complex $s$ plane on the ``unphysical'' sheets.
When using a parametrization of $\Vcal$ with a set of bare poles,
the re-summation of the Bethe-Salpeter equation
eventually converts them into dressed poles.
To see this, consider a situation where $\Vcal$ contains just a single pole term in a coupled channel setting assuming just $S$-waves:
\begin{equation}
    \Vcal_{ba}(s)= \frac{g_a g_b}{M^2-s} \ .
\end{equation}
Then Eq.~\eqref{eq:BSE} is solved by
\begin{equation}\label{eq:singlepole}
    \Mcal_{ba}(s)= \frac{g_a g_b}{M^2-s-\sum_c g_c^2\,\Chew_c(s)}\, \ ,
\end{equation}
which exhibits poles at all those values of $s$, where
\begin{equation}
    M^2-s-\sum_c g_c^2\,\Chew_c(s)=0 \ .
    \label{eq:singleres}
\end{equation}
Here the self-energies need to be continued analytically to the relevant unphysical sheets. Clearly, the solutions of Eq.~(\ref{eq:singleres}) appear typically in complex locations instead of at $s=M^2$.

When the real part of Eq.~\eqref{eq:Chew} is dropped,
Eq.~\eqref{eq:singlepole} reduces to a Breit-Wigner function with energy dependent widths.
In a further simplification that is often used,
the energy dependence of the self-energy is dropped,
by replacing $\sum_c g_c^2\Chew_c\to iM\Gamma$.
A non-relativistic Breit-Wigner amplitude is obtained by writing,
\begin{equation}
    \frac{1}{M^2-s-iM\Gamma} = \frac{1}{2M}\frac{1}{M-\sqrt{s}-i\Gamma/2} + O(\Delta^2)\,,
\end{equation}
where $\Delta = \sqrt{s}-M$.

The approximations are justified only for very special kinematic situations:
all pertinent thresholds
should be very remote with respect to the pole location---the distance to the respective threshold
should be much larger than the width of the resonance.
As soon as there are various resonances in a partial wave,
simply adding Breit-Wigner functions automatically violates unitarity. In contrast, introducing
a potential of the type
\begin{equation} \label{eq:K.matrix}
    \Vcal_{ba}(s)= \sum_\res\frac{g^\res_a g^\res_b}{M_\res^2-s} + \sum_{n=0}^{n_\mathrm{max}} c_{ba}^{(n)} s^n\,,
\end{equation}
into Eq.~\eqref{eq:BSE} automatically maintains unitarity.
With the second term in the sum, we also added non-pole terms without spoiling any property of the $S$-matrix\footnote{It is not excluded that such a potential generates poles in the complex plane of the first sheet---parameters spaces with this property need to be excluded from the space of allowed solutions.}.
The construction of unitary scattering amplitude with Eq.~\eqref{eq:BSEsol} using the on-shell approximation and potential in the form of Eq.~\eqref{eq:K.matrix} is known as the $K$-matrix approach~\cite{Aitchison:1972ay} (often, however, the real part of Eq.~\eqref{eq:Chew} is dropped, potentially causing unphysical singularities to the system).
Left-hand cuts can be included in Eq.~\eqref{eq:K.matrix} without spoiling unitarity constraints, although this is rarely done in practice. For more sophisticated approaches, see, e.g. Ref.~\cite{Heuser:2024biq}.
%